# SEARCH FOR NUCLEAR REACTIONS IN WATER MOLECULES

V.B. Belyaev[1], A.K. Motovilov[1], M.B. Miller[2], A.V. Sermyagin[2], I.V. Kuznetzov[1], Yu.G. Sobolev[1], A.A. Smolnikov[3], A.A. Klimenko[3], S.B. Osetrov[3], and S.I. Vasiliev[3]

[1] *Laboratory of Theoretical Physics, JINR, 141980, Dubna, Russia*
[2] *Institute of Physical and Technology Problems, 141980, Dubna, Russia*
[3] *Institute for Nuclear Research, RAS, 117312, Moscow, Russia*

*Abstract.* A possibility of molecular-nuclear transitions to occur was recently predicted for some few-atomic systems. Among others, the molecule of ordinary water was shown to be a candidate for this effect due to a presence of ($1^-$, 4.522 MeV)-resonance in the $^{18}$Ne nucleus. The search for traces of nuclear reactions was carried out for condensed and vaporous phases of water, with the use of low-background annihilation spectrometry. The measurements were performed under conventional conditions and under conditions of the Baksan Neutrino Observatory.

A detailed consideration of light nuclei spectra reveals numerous cases when nuclear resonance states coincide in energy with a threshold for their decay via two- or three- body channels. Provided products of the decay produce any chemically bound system (a molecule or a radical), it is reasonable to speak of a degeneracy of these molecular and nuclear quantum states. In other words, quantum states of the molecule should be a mixture of pure molecular and nuclear (resonance) states. Then, a new effect can take place.

If, in general case, a coupling between molecular and nuclear states in a few-atomic system is extremely weak due to a wide Coulomb barrier and a short-range character of nuclear forces, the existence of the narrow near-threshold resonance can alter the situation. Under certain condition, large distances may give a non-negligible contribution to an overlap integral of individual wave functions corresponding to pure nuclear or molecular states of the real few-atomic system. Hence, an effect of molecular-nuclear transition (MNT) is expectable in the compounds of this type [1–3].

The most interesting example of such systems is a water molecule $H_2O$ in an excited rotation state $1^-$ which is degenerate with a resonance state (4.522, 1⁻) of $^{18}$Ne nucleus. Fig.1 illustrates this situation, displaying a fragment of level diagram of the $^{18}$Ne [4].

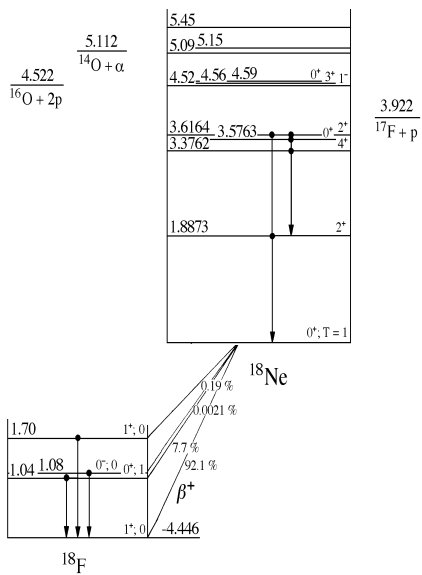

Fig.1. Fragment of $^{18}$Ne nuclear level scheme.

For the first experimental study in this direction the above system $H_2O$ (=2p+$^{16}$O) → $^{18}$Ne →$^{18}$F →$^{18}$O was chosen. An experimental approach was designed taking into account specific molecular-spectroscopy properties of water. Rotational states can be excited only in free water molecules, while their population in the condensed state is prohibited due to the powerful *hydrogen bonding* between the molecules [5]. This gives an opportunity to control the suggested MNT process by means of interchanging conditions for accumulation or decay of the radioactive products of MNT. An over-simplified scheme of the experiment is shown in Fig. 2.

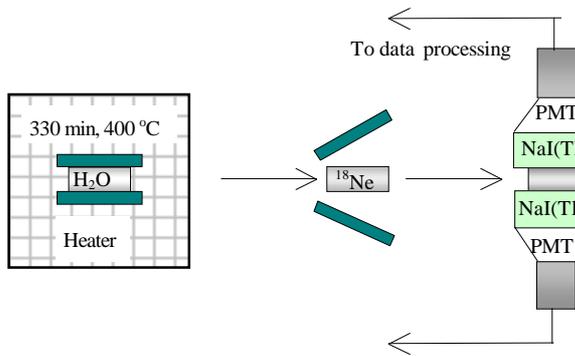

Fig.2. Scheme of the experimental procedure.

During the periods of accumulation, a water sample within a sealed out measuring chamber was heated up to a critical point 647 K, at which temperature the entire amount of water was safely in a vapor phase regardless of pressure. To withstand rather high pressure developed (~22.5 MPa), the chamber (a shortened stainless or titanium cylinder with thin membranes, transparent for the expected radiation, on the faces) was strengthened for the heating periods by thick removable caps on its faces. For the measuring periods the caps were removed, and after due cooling, the chamber was placed between two large NaI(Tl)-scincillators, operating in a coincidence mode.

The experiments were carried out at the Baksan Neutrino Observatory of the Institute for Nuclear Research of the Russian Academy of Sciences. Many efforts were made to subdue the background radiation as much as possible. The cosmic muons were subdued due to power shielding, as the measuring room was located in an inside gallery within the mountain *Andyrchi* near the *Elbrus* where a rock-layer thickness amounted to ~600 m of water equivalent. Interior walls, ceiling, and floor were constructed of special uranium-free concrete, while the scintillators were kept deep underground for a long time, thus short-lived cosmic-ray-induced activities were reduced to minimum.

Annihilation γ-emission with $E_\gamma$=511-keV and a half-life ~110 min was searched for as a signature of a presence of the accumulated nuclide $^{18}$F in the measured sample, – a daughter product of the suggested decay $H_2O$→$^{18}$Ne*(4.522, 1⁻)→$^{18}$Ne$_{g.s.}$(β⁺, 1.67 s)→ $^{18}$F (β⁺, 109.6 min) →$^{18}$O.

For each γγ-coincidence event the time was registered with an accuracy of 1.0 s, i.e., a resolution was quite sufficient for detailed analysis of a temporal dynamics in the range of interest in this experiment. The accumulation-measuring runs were carried out repeatedly, with the accumulation interval of about three half-lives of $^{18}$F, i.e., ~5.5 hours. Time-schedule for these runs is shown in the insert in Fig.3. All initial intervals (the first ones after the end of heating) of every run (of about four half-lives of $^{18}$F, i.e. ~440 min each), were considered as time-periods contributing to the counting in the region of $E_\gamma$=511 keV owing to the decay of the hypothetically accumulated $^{18}$F. The rest parts of measuring cycles were used to estimate a background in the same energy range.

In the Fig.3, spectra of the γγ-coincidences is displayed, which were accumulated in two time intervals followed each other and corresponded to the initial and subsequent three half-life periods of $^{18}$F. The spectra were summed over the nine measuring runs (the run # 2 was too short and was not included to the spectra in Fig.3). For the comparison a calibration spectrum, measured with the $^{22}$Na source is displayed at the same picture.

The coarse-pitch time distribution of the registered events centered on $E_\gamma$=511 keV is shown in the Table 1. It is featured by some time non-stationarity of the process that can be considered as corresponding to the decay with the half-life of the sought after nuclide $^{18}$F, – the intermediary product of the suggested decay chain: $H_2O \rightarrow {}^{18}Ne \rightarrow {}^{18}F \rightarrow {}^{18}O$. The analysis of the data results in a value $T_{1/2} \sim 10^{19}$ years for the decay of water through the studied process of MNT at the confidence level 68% (unless some non-stationary background was left out of account). It is necessary to note that the above value are related to the specific condition the water was under in the experiment (the phase of state, temperature, pressure, etc.).

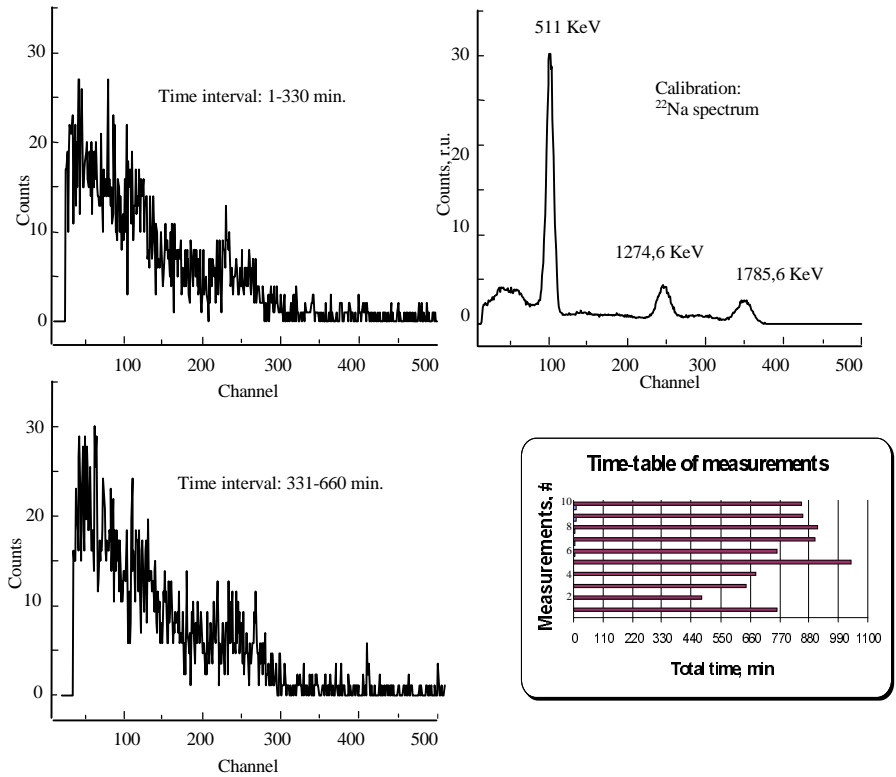

Fig.3. Spectra of γγ-coincidences in the first (upper left) and second (below) time-intervals. To the right: the $^{24}$Na spectrum. At the insert there is a time diagram of measurement runs.

| | Measuring interval | Number of γγ-coincidences ($E_\gamma$=511 keV) | |
|---|---|---|---|
| # | [begin, end], min | Total | Minus background, CL 68% |
| 1. | [1, 110] | 50±7.1 | 10,6±7.9 |
| 2. | [111, 220] | 49±7.0 | 9,6±7.8 |
| 3. | [221, 330] | 46±6.8 | ≤11.7 |
| 4. | [331, 440] | 39±6.2 | ≤6.7 |
| 5. | >440 | *39.4±3.5 | |
| *Notes: The data was processed with the use of Bayes algorithm modified to account statistics of the background [6].* | | | |
| *\*Normalized to 110-min interval; \*\*summed over 1-4 time intervals.* | | | |

With **27±19 summed across intervals 1-4 in the "Minus background" column.

Table 1. "Gross" time distribution of γγ-coincidences at $E_\gamma$=511 keV

A yield of MNT in the condensed state of water was measured also. This part of the experiments was carried out in the conventional ("elevated") lab at the Institute of Physical and Technology Problems (Dubna). In this case, the background was measured with the sample containing heavy water (D$_2$O, 99.0%-enreachment). Thus, the background was measured under the conditions (registration geometry, water amount, γ-ray absorption, etc.)

exactly identical to those under which the effect in $H_2O$ sample was measured. For the value of the half-life of water in condensed state in respect to the $H_2O \rightarrow {}^{18}Ne$ decay a lower limit $\sim 4 \cdot 10^{21}$ years was estimated (99% confidence level).

The total statistics was far from sufficient to allow a decisive conclusion; nevertheless, the results obtained show expediency of further experiments in this line, as well as some theoretical analysis concerning a probability estimate for various systems.